\documentclass[review]{article}

\usepackage{lineno}
\modulolinenumbers[5]
\usepackage[utf8]{inputenc}
\usepackage[T1]{fontenc}
\usepackage{graphicx}
\usepackage{amsmath}
\usepackage{hyperref}
\usepackage{todonotes}

\begin{document}

\begin{flushleft}
{\Large
\textbf\newline{Smart energy management as a means towards \\improved energy efficiency}
}
\newline
\\
Dylan te Lindert\textsuperscript{1},
Claudio Rebelo de S\'{a}\textsuperscript{1,2,*},
Carlos Soares\textsuperscript{3},
Arno~J.~Knobbe\textsuperscript{2}
\\
\bigskip
\bf{1} INESC TEC, Porto, Portugal
\\
\bf{2} LIACS, Universiteit Leiden, Netherlands
\\
\bf{3} Faculdade de Engenharia, Universidade do Porto, Portugal
\\
\bigskip
* c.f.de.sa@liacs.leidenuniv.nl

\end{flushleft}

\begin{abstract}
The costs associated with refrigerator equipment often represent more than half of the total energy costs in supermarkets .
This presents a good motivation for running these systems efficiently.
In this study, we investigate different ways to construct a reference behavior, which can serve as a baseline for judging the performance of energy consumption.
We used 3 distinct learning models: Multiple Linear Regression, Random Forests, and Artificial Neural Networks.
During our experiments we used a variation of the sliding window method in combination with learning curves.
We applied this approach on five different supermarkets, across Portugal.
We are able to create baselines using off-the-shelf data mining techniques.
Moreover, we found a way to create them based on short term historical data.
We believe that our research will serve as a base for future studies, for which we provide interesting directions.
\end{abstract}

\section{Introduction}
Compared to other buildings, supermarkets consume proportionately more energy~\cite{TIMMA2016435,Mylona}.
This is mainly due to refrigeration needed to slow down the deterioration of food, by retaining them on a predetermined temperature~\cite{Mylona}.
Electricity costs associated with refrigeration accounts for a large part of the operating costs because these machines are continually utilizing energy, day and night.
As a result, costs associated with refrigerator equipment can represent more than 50\% of the total energy costs~\cite{Opti,Mavro,FoodRetail,TIMMA2016435}.
Retailers operate in an industry that is characterized as 
competitive and low-margin~\cite{Opti}.
If they are able to become more energy efficient this can make them more competitive.
This outlines the importance of operating the system at its optimum performance level so the associated energy costs can be reduced.

Energy baselining makes it possible to analyze the energy consumption by comparing it to a reference behavior~\cite{Stulka}.
Furthermore, it can be used to measure the effectiveness of energy efficiency policies by monitoring energy usage over time.
Changes in energy policies, such as retrofitting the equipment, can require high investments.
This makes it important for a retailer to know if the investments are truly effective, in the reduction of energy consumption.
To estimate energy savings with reasonable accuracy, the energy baselines need to be accurate. 
It can be challenging to estimate the quality of these energy baselines.
One way is to run the old policies in parallel with the new ones, which is often impossible.
Determining the quality of these baselines can yield significant results for supermarkets.

The objective of this work is to develop energy baselines using off-the-shelf data science technologies.
Different technologies will be tested and applied on the data obtained from several supermarkets to test their performance.
Fives supermarkets in Portugal, will be analyzed as a case study with a methodology based on energy baselining.


\section{Background}
The characteristics of the food-retail industry, such as fierce competition and low margins, makes retailers continually search for ways to operate more efficiently~\cite{Opti}.
Since energy costs are the second highest costs for a retailer~\cite{HPAC}, a decent energy management process is vital for improving efficiency~\cite{SCHULZE20163692}.

Energy Management (EM) has been the subject of numerous studies throughout the years, and, because the field of EM is wide, it can be described in many different ways~\cite{SCHULZE20163692}.
A purpose of EM is to search for improved strategies to consume energy in a more efficient way.
From a business point of view, greater energy efficiency is of importance because it provides a number of direct, and indirect, economic benefits~\cite{WORRELL20031081}.

Several reasons can keep companies from investing in energy efficiency measures ~\cite{Gillingham}.
For example, when inadequate information is available about the results of these investments,
this can limit companies to invest in them~\cite{Gillingham}.
Energy management can focus on addressing these factors to enable businesses to invest.
In order to evaluate the efficiency an energy efficiency measure the observed energy consumption of the store/system must be compared to a \emph{reference behavior}~\cite{Stulka}.
One way to create this reference behavior is to use energy baselining, 
here the reference behavior is defined as the previous, historically best, or ideal, theoretical performance of the given store~\cite{Mavro}.
Energy baselines are usually created on the analysis of historical data~\cite{Stulka} and can be developed using traditional data mining techniques.

Time-series prediction is a method of forecasting future values based on historical data~\cite{CHOU2016751}
In time series forecasting, forecasts are made on the basis of data comprising one or more time series~\cite{chatfield2000time}.
Time series data are defined as the sort of data that is captured over a period of time~\cite{hamilton1994time} (Eq.~\ref{eq:TimeseriesFormula}).
\begin{equation}
\label{eq:TimeseriesFormula}
X_{1},X_{2},\ldots X_{t-1},X_{t}\ldots
\end{equation}
Where $X$ is the value measured at time $t$.
Creating energy forecasts is an important aspect of the energy management of buildings~\cite{WANG2017796}.
Finally, making forecasts can also help in model evaluation when testing different time series algorithms~\cite{chatfield2000time}.

We want to be able to use domain-specific knowledge to engineer new features, therefore, we decided to follow a regression approach.
Regression is not a time series specific algorithm for forecasting, however, it can be applied to make time series forecasts.
In multiple regression models, we forecast the dependent variable using a linear combination of the independent variables.
Based on this relationship the algorithm will be able to predict a value for the dependent variable.

We selected off-the-shelf machine learning algorithms like Multiple Linear Regression (MLR), Random Forests (RF) and Artificial Neural Networks (ANN) to perform the regression.
One way to test the accuracy of the algorithms, is to compare the predicted values with the actual observed values.

\begin{figure}[ht!]
    \centering
    \includegraphics[width=\textwidth]{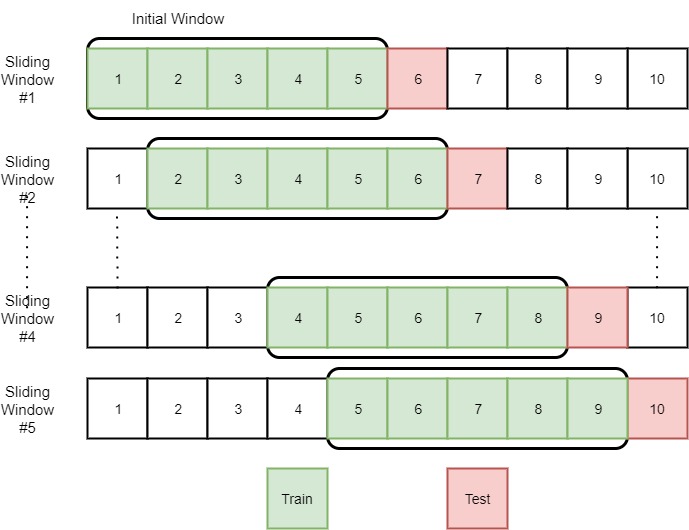}
    \caption{Example of a Sliding Window Validation}    
    \label{fig:TS}
\end{figure}

Nowadays, Machine Learning models and methods are applied in various areas and are used to make important decisions which can have far-reaching consequences~\cite{BERGMEIR2012192}.
Therefore, it is important to evaluate their performance.
Currently, Cross-Validation (CV) is the widely accepted and most used evaluating technique in data analysis and machine learning~\cite{JIANG2017219,BERGMEIR2012192}.
However, Cross Validation does not work well in evaluating the predictive performance of time series~\cite{JIANG2017219}.
One way to validate the prediction performance of a time series model is to make use of a Sliding Window design~\cite{HOOT2008116}, (Figure~\ref{fig:TS}).
In this method, the algorithm is trained and tested in different periods of time.
To evaluate the prediction performance of the algorithms we used the Mean Absolute Error (MAE) as the error metric because the MAE is the most natural measure of the average prediction error~\cite{MAE,Wilmott}.
The following formula shows how the Mean Absolute Error is calculated:
\begin{equation} \label{eq:MAE}
MAE = \frac{1}{N}\sum\limits_{{i = 1}}^N {\left| {\hat{Y_i} - {{Y_i}}} \right|}
\end{equation}
Here \(\hat{Y_i}\) is the predicted value and \({Y_i}\) is the observed value.

Numerous studies focused on energy prediction because forecasting the energy consumption is an important component of any energy management system~\cite{Nasr}.
In New Zealand~\cite{SAFA2017107}, researchers used MLR to calculate the optimal energy usage level for office buildings, based on monthly outside temperatures and numbers of full-time employees.
With this knowledge, they could build an energy monitoring and auditing system for the optimization and reduction of energy consumption.
In the UK~\cite{SPSS}, researchers used an MLR to forecast the expected effect of climate change on the energy consumption of a specific supermarket.
They estimated that, until 2040, the gas consumption will increase 28\%, which is more, compared to the electricity usage, which will increase 5,5\%.

In the UK, most supermarkets negotiate energy prices and, when they exceed their predicted demand, they have to pay a penalty. 
Therefore, their ability to accurately predict energy consumption will facilitate their negotiations on electricity tariffs with suppliers.
One supermarket in the UK used ANN's to analyze the Store’s Total Electricity Consumption as well as their individual systems, such as Refrigeration and Lighting~\cite{Mavro}.
For each of these systems, they developed a model to provide an energy baseline.
This baseline is used for performance monitoring which is vital to ensure systems to perform adequately and guarantee operating costs and energy use are kept to a minimum.
Finally, ANN's have been used for energy prediction with the final goal of estimating the supermarkets future CO2 emissions~\cite{Chari}.

A recent paper~\cite{WANG2017796}, provides a detailed literature review on the state-of-the-art developments of Artificial Intelligent (AI) based models for building energy use prediction.
It provides insight into ensemble learning, which combines multiple AI-based models to improve prediction accuracy.
The paper concludes that ensemble methods have the best prediction accuracy but that a high level of technical knowledge and computational resources is required to develop them.
Consequently, this has hindered their application in real practice.
An advantage of high prediction accuracy is that this can allow early detection of equipment faults that could disrupt store operations~\cite{Mavro}.

These studies show that predicting energy consumption is possible with data mining techniques and that they can predict energy usage within acceptable errors.
Compared to other engineering methods, ensemble methods require less detailed information of the physical building parameters~\cite{WANG2017796}.
This saves money and time in conducting predictions compared to simulation tools.
Hence, they could replace them in the future.
Because studies use different types and volumes of input data, there is no unified input data format.
Therefore, knowledge of the methods and a variety of data is needed to create meaningful and accurate predictions.

\section{Defining baselines with Machine Learning Algorithms}
Every forecast $\hat{Y_i}$ of an observed value ${Y_i}$ will have a forecast error $E$, which describes the deviation among them.
These deviations 
can result from poor prediction performance or energy savings/losses.
It is very hard to forecast a numeric value correctly, the deviations can be larger or smaller.
Thus, to provide good estimates of the effect of changes in energy management policies, it is important to have a learning model that can create energy baselines as accurate as possible.

The objective of this study is to asses the reliability of the learning model in different aspects.
First, we want to determine which model is best in creating a reliable baseline with the least amount of training days.
This can be beneficial in two specific situations: when a retailer opens a new store, or implements new energy policies.
When a new store is opened, no data has been collected about the energy performance of \emph{this} specific store.
To create a baseline as soon as possible, it is essential to know how many days it takes to collect sufficient data.
Therefore, we study the minimum amount of days needed to create a reliable baseline.
This information is also suitable for updating the baseline when the configuration of the store changes, e.g., due to upgrades of the refrigeration equipment.

When we know this setup, we want to discover the lifespan of this prediction, i.e., how long does this energy baseline remain reliable after being learned.
It is important to determine how reliable the baseline is and if it needs updating, because we expect that the prediction error will grow over time.
As a result, the prediction error will behave differently for short and long term predictions.
With this information, the life-cycle of a model can be determined, which defines how often the model needs to be updated.

When a new energy saving policy is implemented, the Retailer wants to estimate how much energy is saved.
Therefore, a model has to be developed which is able to make long term predictions based on the old configuration of the store.
With this baseline, the Retailer can see what the estimated energy consumption would be if they did not change the layout.
By comparing this baseline with the observed energy consumption or the new baseline, the difference can be estimated.
We will examine the behavior of the model for long term predictions because the Retailer needs to know for how long he can estimate, with a reasonable accuracy, the energy gains from a certain energy policy.

\subsection{Approach}
We obtained time series data from five supermarkets across Portugal, which consist of  measurements of the \emph{Refrigeration Energy Consumption}, \emph{Outside temperature} and the \emph{Timestamp}.
The original time series data was provided, in sometimes irregular, 15-minute intervals.
After this restructuring, the data is converted into hourly values and eventually, transformed to daily formats.
The energy consumption is measured in kilowatt hour (kWh) from the Retailer's energy monitoring system.
The weather data consists of the outside temperature derived from a sensor placed on the roof of the store and is measured in degrees Celsius (C$^{\circ}$).

In order to apply a similar approach to the data of each store, we decided to work separately with datasets that have a similar structure.
We will use domain knowledge to create features for the datasets.
The process of designing new features, based on domain knowledge, is called Feature engineering~\cite{LI2017232}.
Before creating these datasets, we first identified the dependent and the independent variables. 
In this study, an energy baseline will be created that reflects the estimated refrigeration energy consumption.
Consequently, this will be the dependent variable, and the independent variables are the ones influencing this consumption.
Only the factors that are measured, by all stores, can be used here as an independent variable.


\subsection{Estimating Reliability}
For a retailer it is important to estimate, with reasonable accuracy, the energy savings resulting from energy policies.
If we train an algorithm with data before a energy policy change, we can create an energy baseline that shows what the energy consumption would be if this policy has not been changed.
By comparing this energy baseline with the observed consumption, after the policy change, we can estimate the energy savings.

The first objective of this study is to define the minimal set of training examples needed to build a reliable energy baseline.
To do this, we train the machine learning algorithms with different numbers of training days.
Each iteration we increase the number of training examples and evaluated the models’ prediction accuracy. 
When all iterations have been completed, we are ready to plot the error metrics in the learning curves.
Because this approach is replicated for the three algorithms, this also reveals which one performs best.

After we selected the learning model which is able to create the baseline with the least amount of data, we define the update frequency of this setup.
We expect the prediction error to grow over time, and therefore the energy baseline will become unreliable at some point when the prediction error becomes too high.
To find the point of which we recommend updating, we use the previously defined setup, to make predictions for the remaining dataset.
As soon as the predictions are made, we compute a MAE for each of 10 subsequent predictions.
Once all the errors are computed, we can plot them to see how the prediction error develops over time.
This enables us to analyze how the prediction accuracy develops along the prediction horizon, and define the update frequency.

Finally, the third part of this research is to analyze the long term prediction performance.
This was done by training each model with various sizes of training data and let it predict for the remaining dataset.
After the predictions were made, we then calculated a MAE for every 10 subsequent predictions.
Having plotted the error metrics meant that we could study their performance over time.

\section{Experimental Setup}
In order to study the three objectives described before, we designed an approach based on Learning curves in combination with Sliding windows.
Our experimental setup is a variation of the Time series approach used by~\cite{Busetti,vanRijn2015}.
The method we propose is visualized in Figure~\ref{fig:SDLC}.
We decided to use this particular method because we want to train machine learning models with different sizes of historical training data.
The learning curves enable us to visualize and evaluate their performance.

\subsection{Data}
The studied datasets are mainly based on the energy consumption and weather data for the whole year of 2016 and the first half of 2017 (Table~\ref{tab:OriginalData}).
The data for each store is available from the moment the store opened or started to collect the data.
Hence, for each store, the maximum amount of data is available.

\begin{table}[hpt]
\caption{Overview Datasets}
\label{tab:OriginalData}
\centering
\small
\renewcommand{\arraystretch}{1.25}
\begin{tabular}{llll}
\hline \hline
Store & First day & Last day & Observations \\ 
\hline
Aveiro               & 04/12/2015 & 26/04/2017 & 510 days   \\  %
Fatima              & 07/01/2016 & 26/04/2017 & 476 days   \\ %
Macedo de Cavaleiros	 & 13/11/2015 & 26/04/2017 & 531 days   \\  %
Mangualde               & 16/05/2016 & 16/05/2017 & 366 days  \\  %
Regua              & 16/05/2016 & 16/05/2017 & 366 days  \\   
\hline \hline
\end{tabular}
\normalsize
\end{table}

Based on the two available variables, \emph{Timestamp} and \emph{Outside temperature}, we created new features with additional information that the algorithm can use.
Designing appropriate features is one of the most important steps to create good predictions because they can highly influence the results that will be achieved with the learning model~\cite{SILVA2014395}.
To determine which features to create, knowledge about the behavior of the store is important~\cite{Mavro}.
The domain knowledge required for this process, was acquired through conversations with experts, reviewing similar studies~\cite{Mavro,SPSS,SAFA2017107,Chari,KARATASOU2006949,Jacob,OROSA201289} and using descriptive data mining techniques, e.g., Subgroup Discovery (SD).
SD is a method to identify, unusual, behaviors between dependent and independent variables in the data~\cite{WIDM1144,Herrera2011}.
In this study, SD will be used to improve our understanding of the behavior of the energy consumption.
Table~\ref{tab:Features} gives an overview of the created features.
\begin{table}[hpt]
\caption{Overview Features}
\label{tab:Features}
\centering
\tiny
\renewcommand{\arraystretch}{1.25}
\begin{tabular}{llll}
\hline \hline
Name & Type & Description & Derived from \\ 
\hline
Weekday           & Categorical (1-7)  & Day of the week    & Timestamp \\ 
Week of the Month & Categorical (1-4)  & Week of the Month  & Timestamp \\
Workday           & Binary (0-1)    & Workday or Weekend & Timestamp \\

Max Temperature  & Numerical & Max Temperature of the Day      & Temperature \\
Mean Temperature & Numerical & Mean Temperature of the Day     & Temperature \\ 
Min Temperature  & Numerical & Min Temperature of the Day      & Temperature \\ 
Temperature Amplitude & Numerical & Absolute Difference Min and Max & Temperature \\

Max Temperature Y..  & Numerical & Max Temperature of Yesterday    & Temperature \\ 
Mean Temperature Y.. & Numerical & Mean Temperature of Yesterday   & Temperature \\ 
Min Temperature Y..  & Numerical & Min Temperature of Yesterday    & Temperature \\ 
Temperature Amplitude Y..  & Numerical & Absolute Difference Min and Max & Temperature \\
\hline \hline
\end{tabular}
\normalsize
\end{table}

\subsection{Algorithms}
We selected off-the-shelf machine learning algorithms like Multiple Linear Regression (MLR), Random Forests (RF) and Artificial Neural Networks (ANN) to perform the regression.

Linear regression is a simple and widely used statistical technique for predictive modeling~\cite{SPSS}.
It has been used before to predict the future energy consumption of a supermarket in the UK~\cite{SPSS}.
The RF is considered to be one of the most accurate general-purpose learning techniques available and is popular because of its good off-the-shelf performance~\cite{Fernandez,Biau}.
Finally, Artificial Neural Networks have successfully been used in recent studies to predict energy consumption~\cite{Mavro,Chari,WANG2017796,KARATASOU2006949,Nasr,FOUCQUIER2013272}.


\subsection{Performance Estimation}

\begin{figure}[ht!]
    \centering
    \includegraphics[scale=0.6]{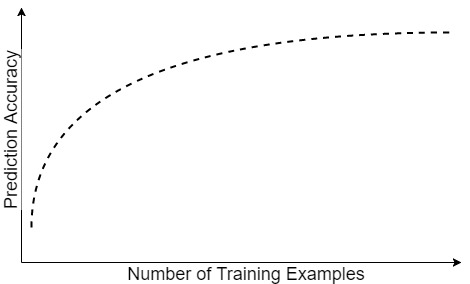}
    \caption{A graphical representation of a learning curve}
    \label{fig:LC}
\end{figure}

In Machine Learning, learning curves are used to reflect the predictive performance as a function of the number of training examples~\cite{LC}.
Figure~\ref{fig:LC} reveals the developing learning ability of a model when the number of training examples increases.
The curve indicates how much better the model gets in predicting when more training examples are used.
The general idea is to find out how good the model can become in predicting and what the subsequent number of training examples is~\cite{LC}.
Since we are searching for the minimum number of training days to create a baseline, we can use the learning curves to identify this number.

To test the learning ability of a model one can create several training sets of data and evaluate their performance on a test set~\cite{Langley1988}.
These training sets can differ in, e.g., volume.
It is preferred that the data for these sets are randomly selected from the available data~\cite{Langley1988}.
The purpose is to train the model multiple times, and after every training, the model performance should be tested.
The results of these tests can be plotted to draw a learning curve which shows the evolution in the performance of the model.
These curves can be clarifying, especially when the performance of multiple models is compared.
Besides for model selection, also the performance of a model can be compared in relation to the number of training examples used~\cite{LC}.
Such a learning curve will tell how the model behaves when it is constructed with varying volumes of training data.

\begin{figure}[ht!]
    \centering
    \includegraphics[scale=0.45]{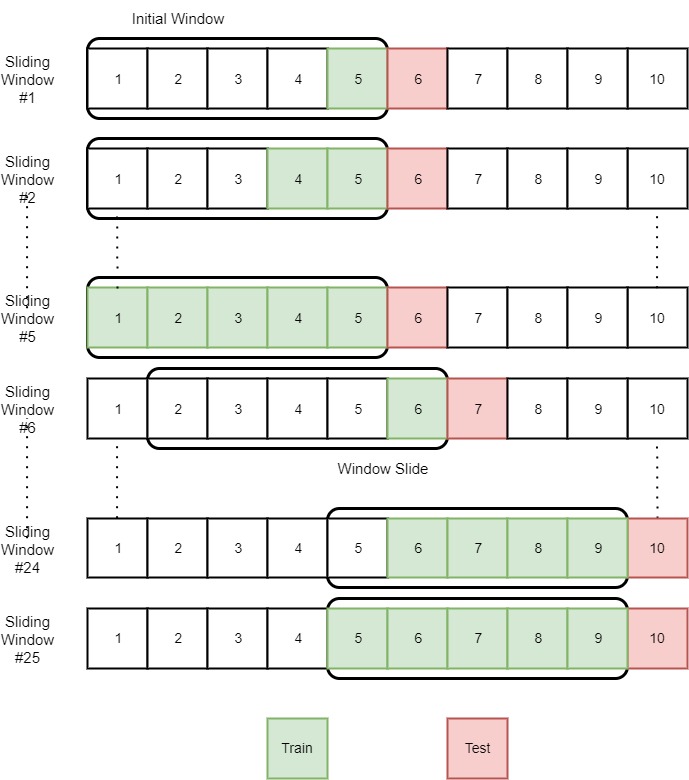}
    \caption{Example of our Sliding Window Process}    
    \label{fig:SDLC}
\end{figure}

\section{Results}

\subsection{Reliability of baselines}
\label{sec:RS2}

\begin{figure}[ht!]
    \centering
    \includegraphics[width=\textwidth]{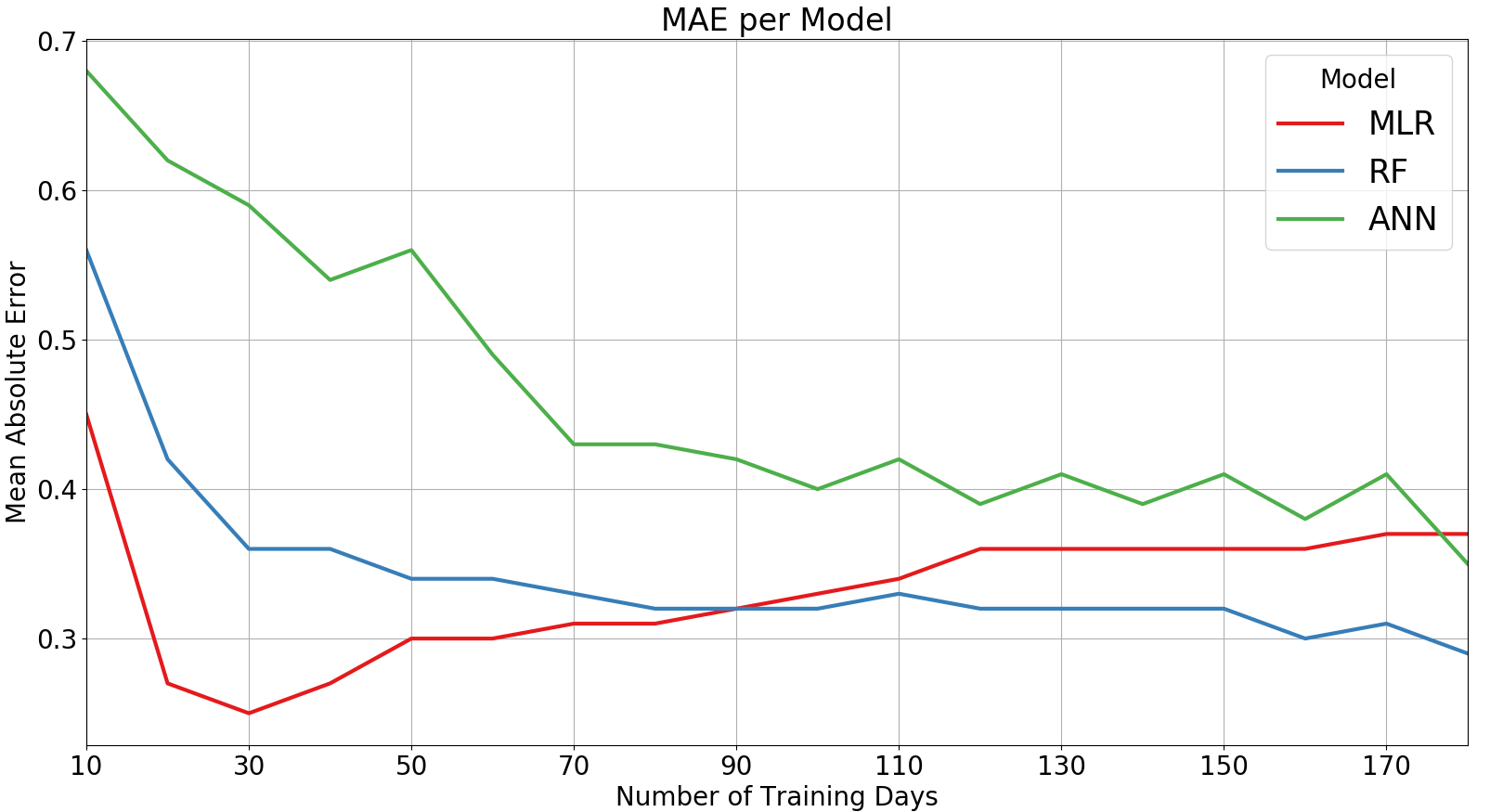}
    \caption{Learning curves, based on an average for all stores and methods}    
    \label{fig:TotalModels}
\end{figure}

In Figure~\ref{fig:TotalModels}, we see how the error evolves as we train the model with more data points i.e.\ days.
This plot displays the learning curves obtained for each of the trained models, MLR, ANN, and RF.
The number of training examples ranged from 10 up to 180 days, with threads of 10, and have been tested for a period of 50 days.
Each line represents the mean of 18 iterations, for all stores, Aveiro, Fatima, and Macedo Cavaleiros, we performed six iterations regarding the method visualized in~\ref{fig:SDLC}.

In Figure~\ref{fig:TotalModels}, we observe that the MLR is the most reliable by a number of 30 days with a MAE of 0.25.
Besides, we observe that using the MLR, as we expand the size of training examples, there is an increase in the MAE.
Furthermore, we perceive a different behavior for the other two learning models.
We see that the performance of the RF stabilizes when we increase the training data following 70 training examples up to 180.
Moreover, we remark that the ANN exhibits a continuous reduction in the MAE when more training examples, up to 180, are attached to the training set.

The learning curves in Figure~\ref{fig:TotalModels}, reveal that each of the learning models is affected differently by the change in the training set size.
We notice that the MLR outperforms the other two methods, for making a reliable baseline using the least amount of days.
Furthermore, we see that the performance of the MLR worsens when we increase the number of training examples.
This can be explained by the nonstationary nature of the datasets.
This non-stationarity is a problem for the MLR since it has difficulties with nonlinear relationships. Because the MLR works well with a smaller number of training examples, we assume that the dataset contains periods of local stationarity.
One study~\cite{LOCALStationary}, shows that it is possible that nonstationary time series appear stationary when examined close up.
In this local period, the statistical properties change slowly over time.
As a consequence, the data that lies close to the forecast period is more likely to be predictive for this forecast period.

For the ANN and RF, stationarity is irrelevant since they are able to handle more complex, nonlinear relations.
We see evidence for this in our results, there is a promising development over time in the associated learning curve.
We believe that with more diverse data, the ANN could be able to predict a baseline with less number of training days than the MLR.
Unfortunately, we were not able to investigate this further.

As shown in Figure~\ref{fig:TotalModels}, we are able to create a reliable model with the MLR trained on 30 days.
Therefore, we trained the MLR for each of the stores during the same period of the year, March 2016, and we estimated the energy consumption for the period of one year, from April 2016, until February 2017.

\begin{figure}[ht!]
    \centering
    \includegraphics[width=\textwidth]{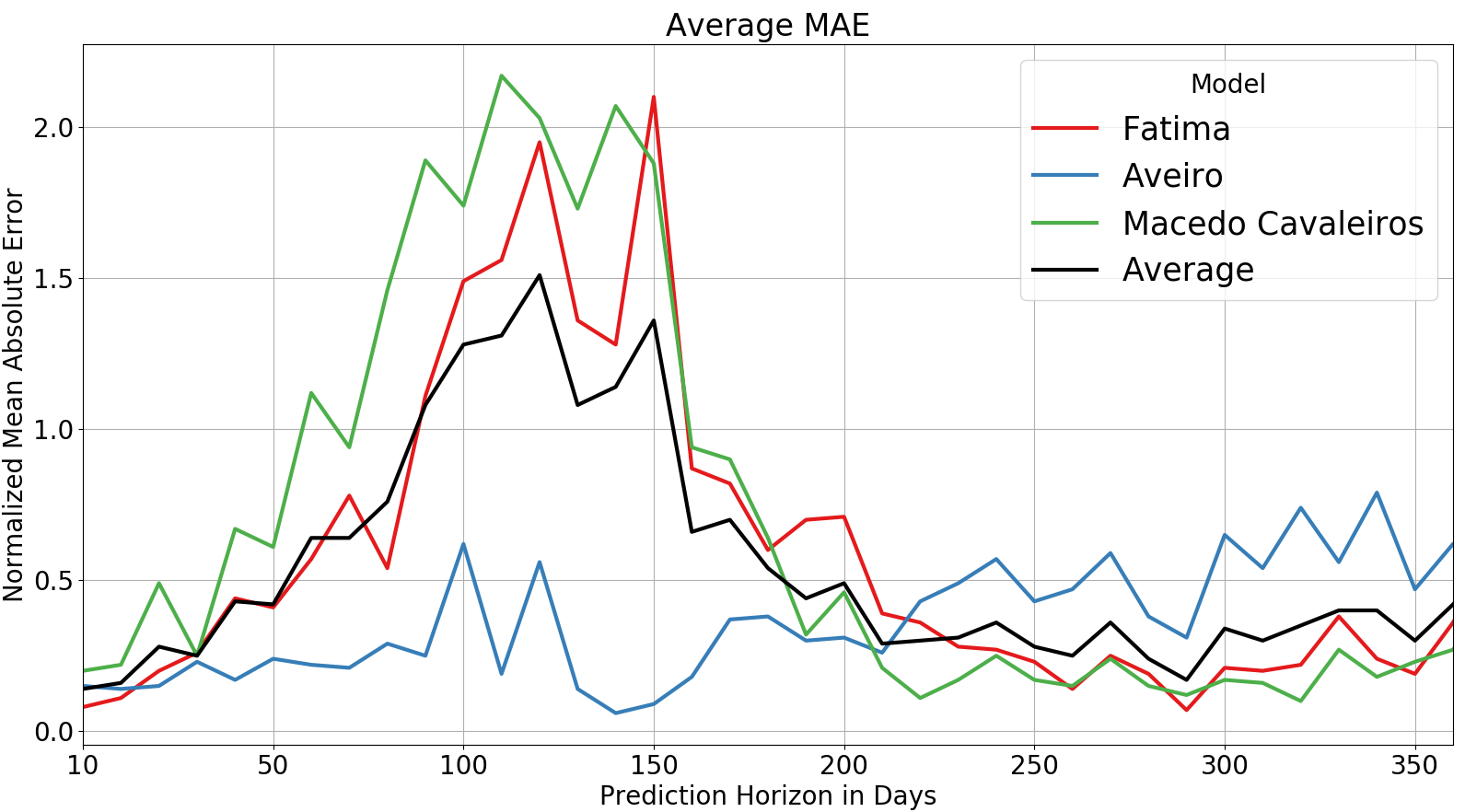}
    \caption{MAE over time using MLR}   
    \label{fig:LC30}
\end{figure}

Figure~\ref{fig:LC30}, shows the evolution of the MAE throughout this period.
We observe that during the first 30 days of predictions, the MAE remains quite low, under 0.5.
Next, we see that during the period between 50 till 180 days, the MAE is higher for all the stores.
As a matter of fact, this period represents the months June, July, August, and September.
Table~\ref{tab:AVGTEMP} shows, that throughout these months, temperature levels reach higher values than in March, the period that was used for training the model.
This explains why the MAE is higher.
To avoid this problem, we could train a different model for each of the two energy profiles. 
Because our dataset is limited, we were not able to test this in practice.

We observe, in Figure~\ref{fig:LC30}, that in Aveiro the influence of seasonality is less evident than for the supermarkets in Fatima and Macedo Cavaleiros.
Since all stores are trained and tested with the same model and in the same period of time, the most plausible factor, for this, are the variables that are related to Temperature.
The average temperatures of the three stores follow a similar pattern, higher in the summer and lower in the winter.
However, if we focus on the amplitudes of the average temperatures per month, (Table~\ref{tab:AVGTEMP}), we observe that Aveiro registered the smallest amplitude, with a difference of $9 C^{\circ}$.
The other stores, Fatima and Macedo Cavaleiros, noted an amplitude of $13 C^{\circ}$ and $18 C^{\circ}$ respectively.
This seems to explain why the model trained for the store of Aveiro, is less affected by seasonality.

In Figure~\ref{fig:LC30}, we notice that after 220 days the accuracy of the model increases again.
When we look at Table~\ref{tab:AVGTEMP}, we see that the temperature values from November on, are comparable to the ones in March.
Nevertheless, the error is still higher than in the period of the first 30 days.
We applied this method in different periods of time, and we perceived similar behavior.

In conclusion, we base our decision on the average prediction.
Figure~\ref{fig:LC30} shows that the average prediction remains stable until 30 days, therefore, we recommend updating the model up to 30 days.

\begin{table}[hpt]
\caption{Average Temperature per Month and Store}
\label{tab:AVGTEMP}
\centering
\small
\setlength\tabcolsep{3.5pt}
\begin{tabular}{lllllllllllllll}
\hline \hline
Store & Jan & Feb & Mar & Apr & May & June & July & Aug & Sep & Oct & Nov & Dec \\ 
\hline 
Aveiro & 12 & 13 & 14  & 16 & 17   & \textbf{20}   & \textbf{21} & \textbf{21} & 19 & 18 & 14 & 14 \\
Fatima & 9  & 11 & 11     & 14 & 15 & 19 & \textbf{22} & \textbf{22} & \textbf{20} & 17 & 12 & 10 \\
M. Cav. & 8  & 10 & 12  & 15 & 16 & \textbf{22} & \textbf{26} & \textbf{25} & \textbf{22} & 16 & 10 & 8 \\ 
\hline \hline
\end{tabular}
\end{table}

\subsection{Estimated energy savings}

Each store has a different number of observations, and they are also collected in different periods of time.
We will train the MLR, RF, and ANN with the first 180 and 360 days of data, and test for the remaining days.
We will do this for the stores located in Aveiro, Fatima, and Macedo Cavaleiros.
Therefore, we train each store in different periods, and not within the same period.

\begin{figure}[ht!]
    \centering
    \includegraphics[width=\textwidth]{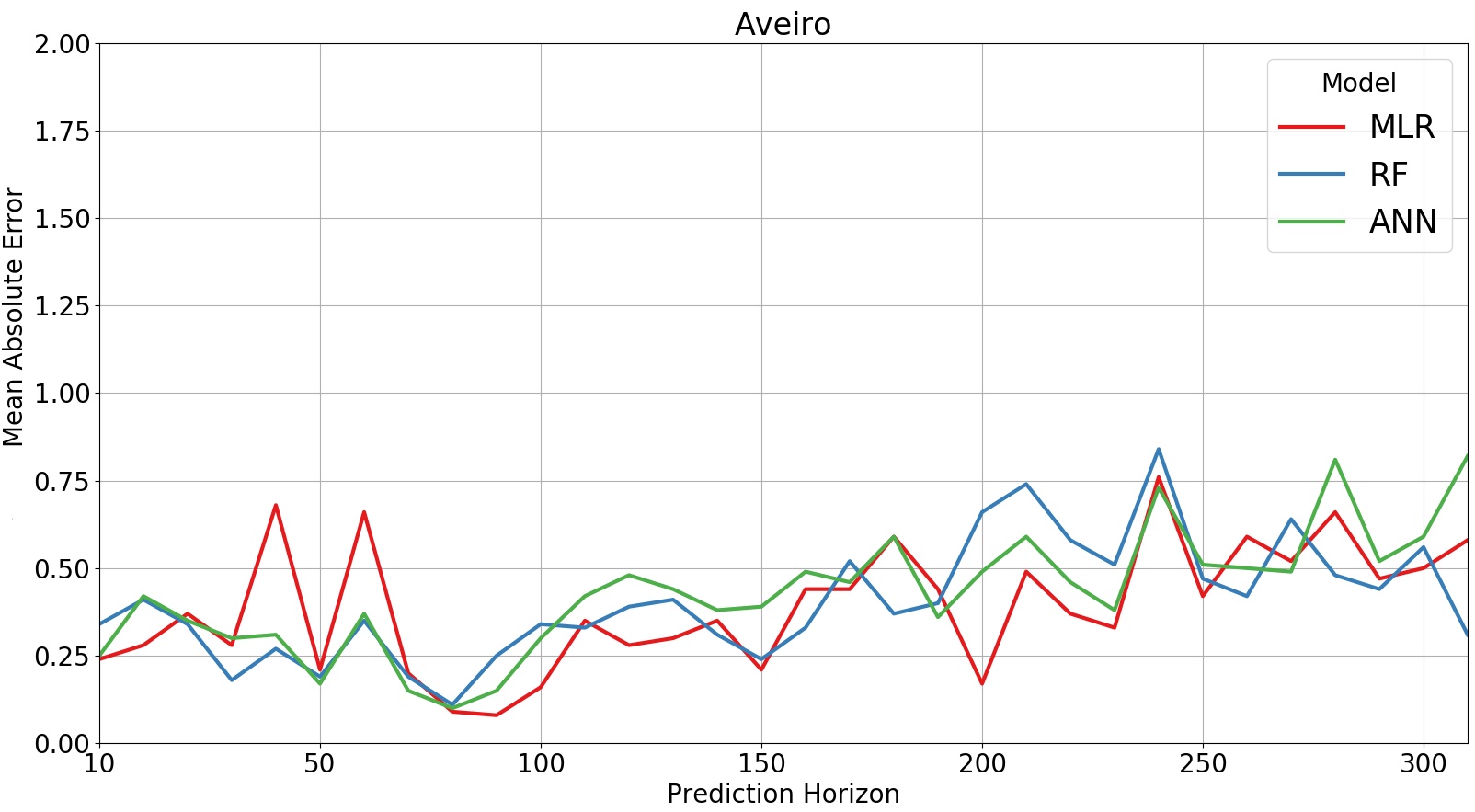}
    \caption{MAE over time using 180 training days, Aveiro}    
    \label{fig:AV180}
\end{figure}

\begin{figure}[ht!]
    \centering
    \includegraphics[width=\textwidth]{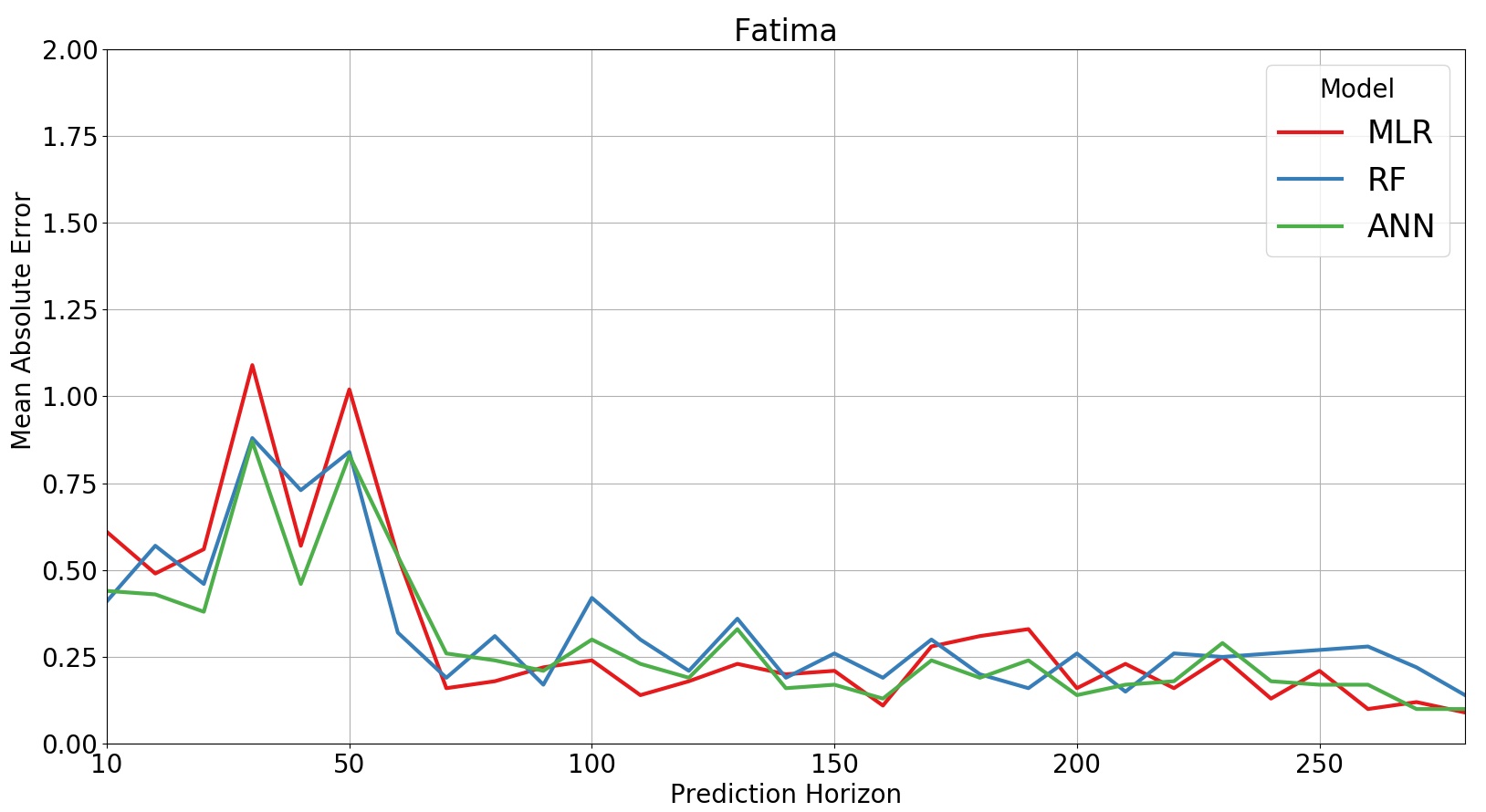}
    \caption{MAE over time using 180 training days, Fatima}    
    \label{fig:FA180}
\end{figure}

\begin{figure}[ht!]
    \centering
    \includegraphics[width=\textwidth]{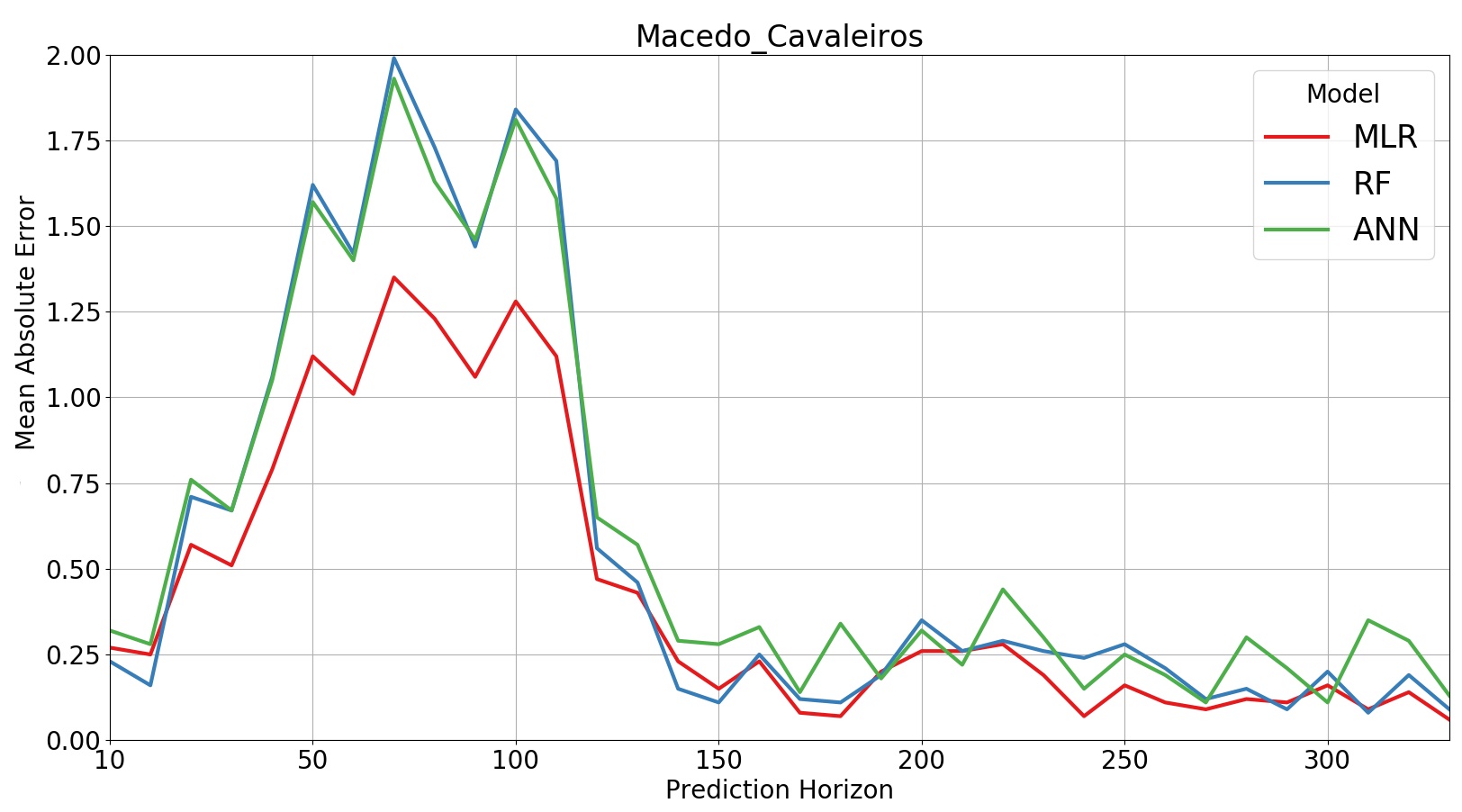}
    \caption{MAE over time using 180 training days, Macedo Cavaleiros}    
    \label{fig:MC180}
\end{figure}

\begin{figure}[ht]
    \centering
    \includegraphics[width=\textwidth]{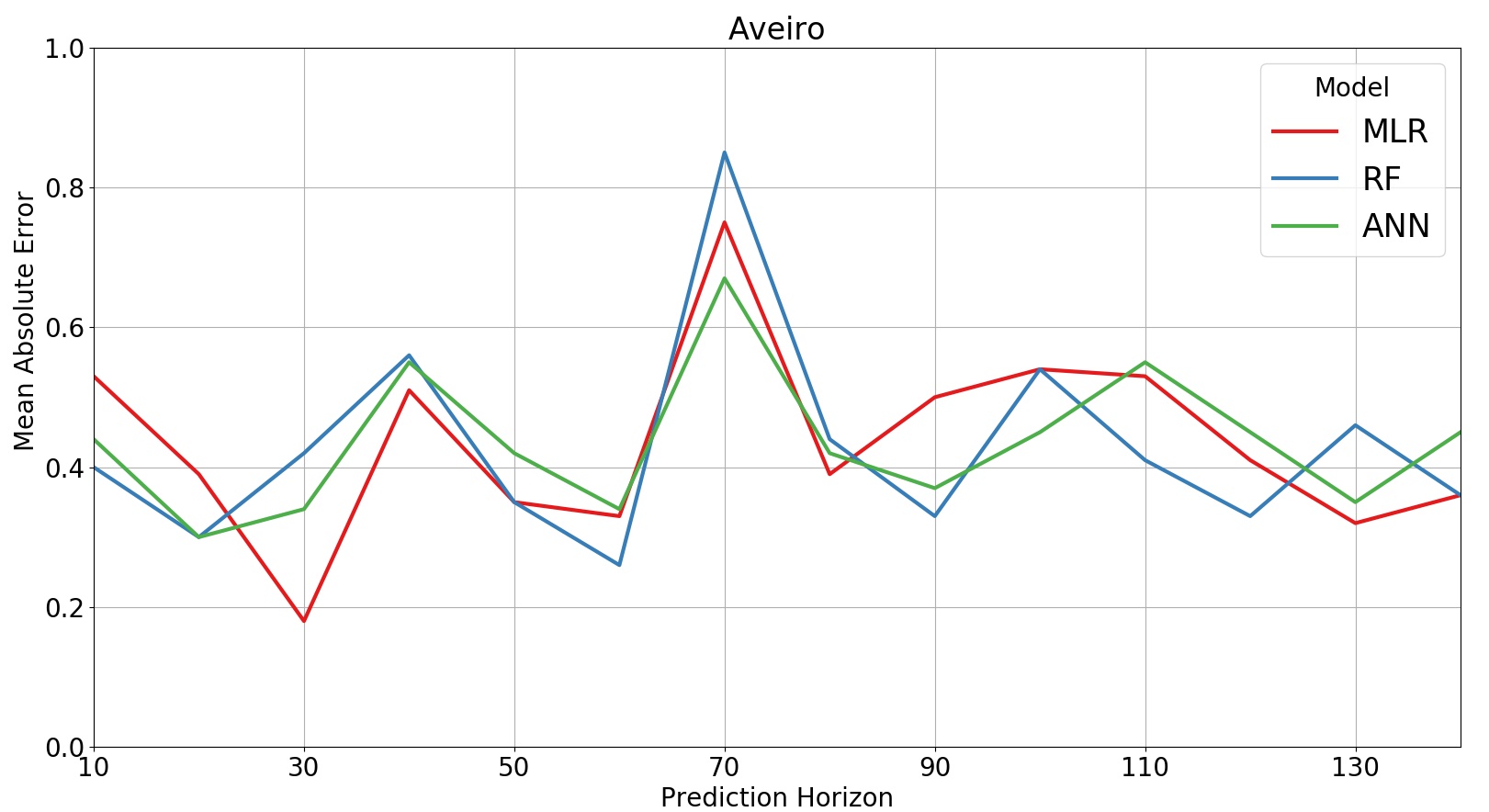}
    \caption{MAE over time using 360 training days, Aveiro}    
    \label{fig:AV360}
\end{figure}

\begin{figure}[ht]
    \centering
    \includegraphics[width=\textwidth]{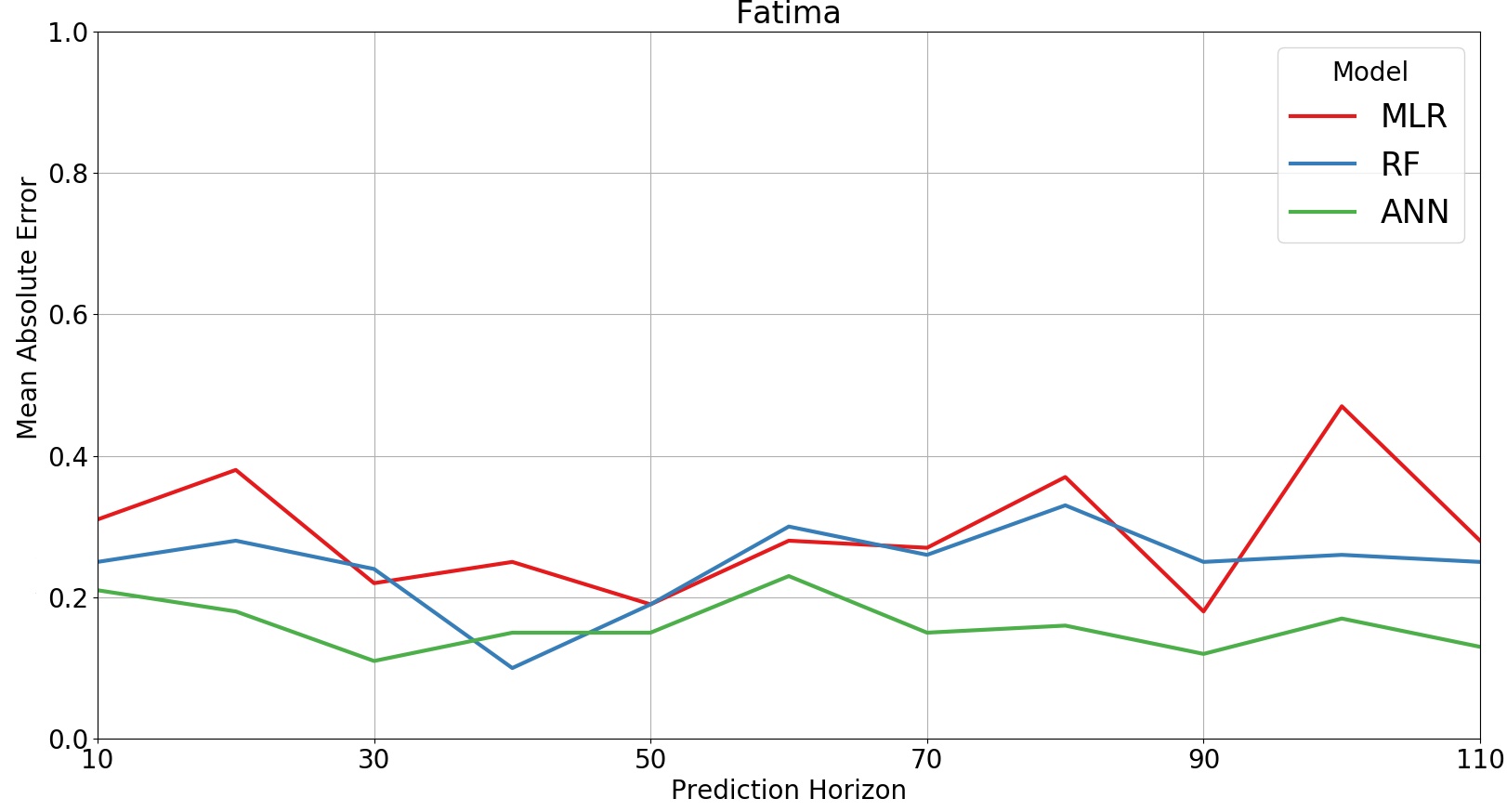}
    \caption{MAE over time using 360 training days, Fatima}    
    \label{fig:FA360}
\end{figure}

\begin{figure}[ht]
    \centering
    \includegraphics[width=\textwidth]{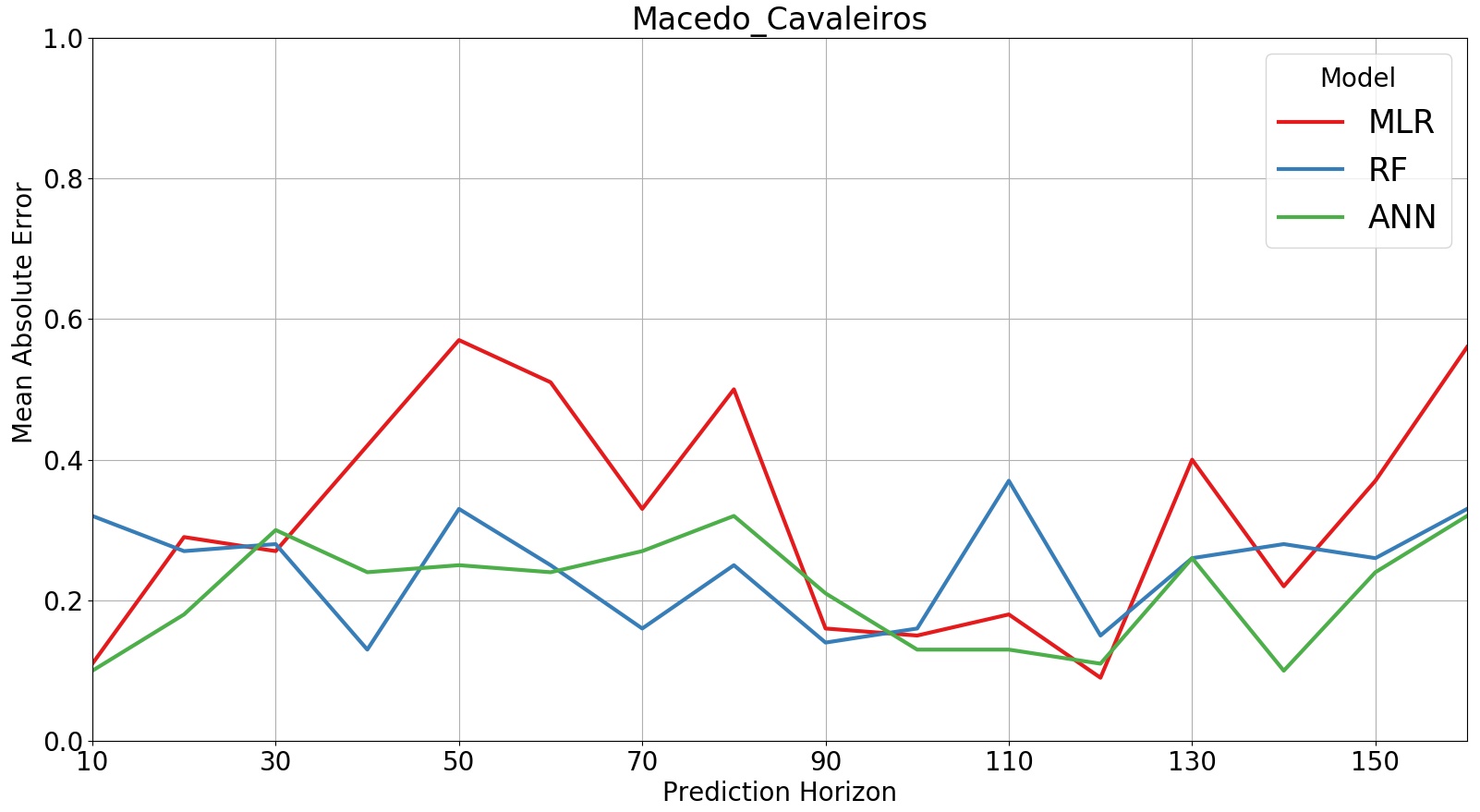}
    \caption{MAE over time using 360 training days, Macedo Cavaleiros}    
    \label{fig:MC360}
\end{figure}

In Figure~\ref{fig:LC30}, we noticed that 30 training days were not enough to make accurate long term predictions.
Therefore, we decide to include more training days into our training set.
Each of the following plots, in Figures~\ref{fig:AV180},~\ref{fig:FA180},~\ref{fig:MC180},~\ref{fig:AV360},~\ref{fig:FA360}, and~\ref{fig:MC360}, show how the prediction error evolves over time, per store, per model and number of training days.
Each point shows the average error for 10 subsequent predictions.

Figures~\ref{fig:AV180},~\ref{fig:FA180}, and~\ref{fig:MC180} show the evolution of the prediction error when the models are trained on the first 180 days of data.
We observe, that each store shows a similar behavior as shown in Figure~\ref{fig:LC30}.
This is more evident when we compare the error of the MLR (red line) with the error in Figure~\ref{fig:LC30}.
Overall, the MAE is lower for the stores of Fatima and Macedo Cavaleiros, if we use 180 days instead of 30 days.
These results also show, that the effect of the different consumption modes is still visible, but less dramatically.

We expect that long term predictions become more accurate when we use 360 training days to train the model because the model is trained with data from all periods of the year.
Because we use this number of training days, a bigger variation of temperature values is included in the training set.
Therefore, we decided to train the models, for all stores, on the first 360 training days and study the predictions on the remaining days.
Figures~\ref{fig:AV360},~\ref{fig:FA360}, and~\ref{fig:MC360} show us how the MAE error evolves for this period of time.
We observe, that the for the corresponding period of time, the MAE is a bit lower than for the models trained on 180 days.

In contrast to Figure~\ref{fig:TotalModels}, the MLR has the worst performance, while the RF and ANN perform somewhat similar.
The results of this experimental part supports the general idea that when we train the models with more data, our predictions will improve.

When the algorithms are trained with 180 training days, the effect of the different energy consumption modes is still visible.
When we use 360 training days, we observe that the predictions become more accurate.
Therefore, we advice to train algorithms on 360 training days to create long term predictions.

\section{Estimate Energy Savings}
\label{sec:Energysavings}
The Retailer wants to estimate, with reasonable accuracy, the energy savings resulting from its energy policies.
Changes in energy policies, such as the retrofitting an equipment, require high investments.
This makes it important for the Retailer to know if the investments are truly effective, in the reduction of energy consumption.
If we use a baseline trained with data before some measure is implemented, we can estimate the energy savings by comparing its estimates with the observed consumption.

We selected two stores that have undergone a retrofitting of the equipment.
From these stores exactly one year of data is available.
Mangualde and Regua had, respectively, 170 and 200 training days available before the Retrofit.
Because we have less than a year of data available, we decide to use the MLR, trained on 30 days, which shows the best performance in Figure~\ref{fig:TotalModels}.

\begin{figure}[hpt!]
    \centering
    \includegraphics[width=\textwidth]{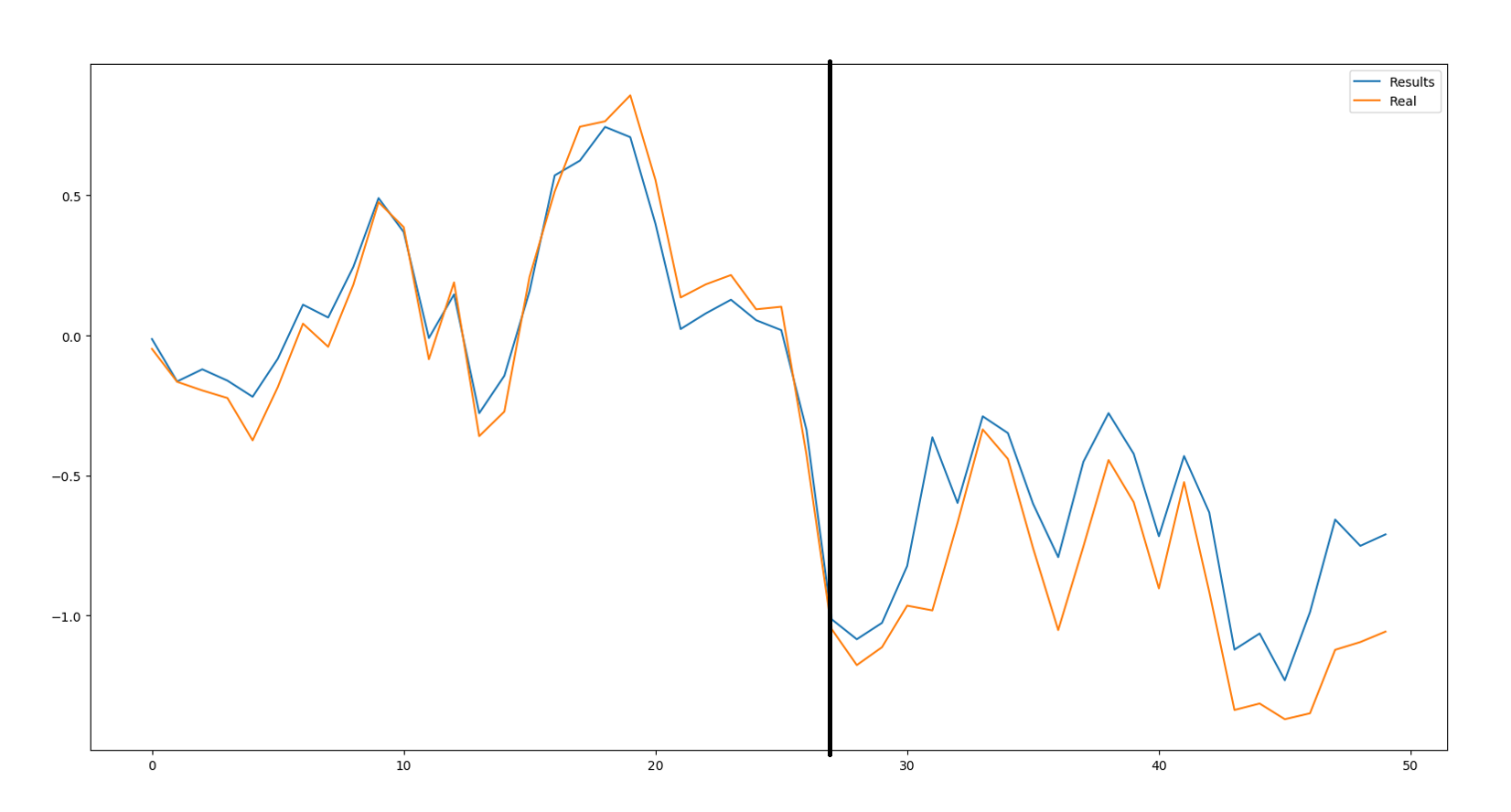}
    \caption{Example of the predicted and observed Energy Consumption, Mangualde}    
    \label{fig:Mangualde}
\end{figure}

\begin{figure}[hpt!]
    \centering
    \includegraphics[width=\textwidth]{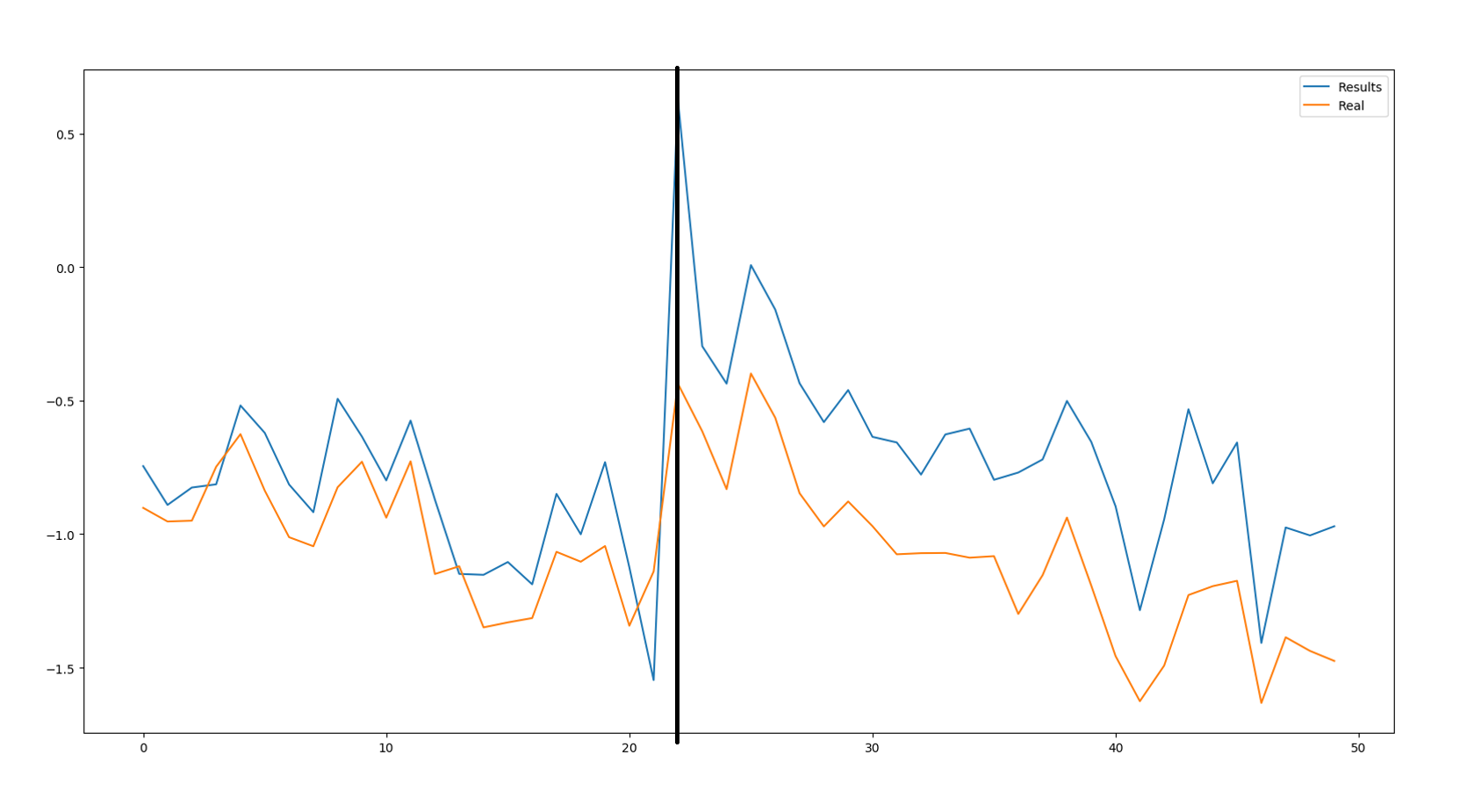}
    \caption{Example of the predicted and observed Energy Consumption, Regua}    
    \label{fig:Regua}
\end{figure}

Figures~\ref{fig:Mangualde} and~\ref{fig:Regua} show the observed consumption (orange lines) versus the baseline estimates (blue lines) for these two stores.
We trained the MLR for both stores, on 30 training days, between 50 and 20 days before the Retrofit and we predicted for 50 days.
This makes it easier to visualize how the baseline compares with the energy consumption before and after the Retrofit.

The deviations, between the baseline and the energy consumption, can result from poor prediction performance or energy savings/losses.
We chose a setup that gives us a reliable baseline, therefore, we believe that the deviations are caused by energy savings.
In both Figures~\ref{fig:Mangualde} and~\ref{fig:Regua}, we observe that, before the Retrofit, the baseline and the real energy consumption intertwine in several points.
This behavior, which was also seen before, shows that the predictions are close to the real consumption.
After the Retrofit, however, the observed consumption is always lower than the prediction, which offers strong evidence that the implemented measure was effective.

Hence, if we assume that the baseline is accurate enough, we can estimate the energy savings using the difference between the predicted and observed energy consumption.

\section{Conclusions}

Energy efficiency measures can require high investments.
This makes it important for the Retailer to know if the investments are truly effective, in reducing energy consumption.
Energy baselines can be used to study the effectiveness of energy efficiency measures.
The results can simplify decisions to reserve funding for the required investments in other stores.

In this study, we researched if off-the-shelf data science technologies can be used to create energy baselines that support improved energy management. Before that, we also performed some exploratory analysis to better understand the data.


Our first goal, was to determine the minimum amount of training days needed to create a reliable baseline, and which model performs best.
For that, we studied the prediction accuracy of three machine learning models, ANN, RF, and MLR, based on various datasets.
For the experiments, we proposed a sliding window approach in which we systematically expanded the size of the training set with historical data.
Our experiments show, that the MLR has a clear advantage over the other two methods for creating a baseline with a minimum amount of days.
This model needs 30 training days to estimate a reliable baseline.

The second goal was to determine how often the algorithm needs to be updated when trained with a MLR on 30 training days.
We trained our algorithm multiple times, on all stores, and in different time periods.
Our analysis shows that the MAE stays low for a period of 30 days, after this the MAE dramatically increases.
Moreover, we observed that the energy consumption follows a different profile when average temperatures are higher than 20 degrees.
These findings are in line with our insights derived from Subgroup Discovery.
Our analysis shows, that the amplitude of the average temperature affects the prediction performance.
Hence, we advise updating the model up to 30 days.

Our third goal, was to determine if we can estimate energy savings after implementing an energy efficiency measure.
To answer this question, we trained our models with 180 and 360 training days and predicted for the remaining days.
Our findings show, that the predictions become the most accurate when trained with 360 training days.
Because we use 360 training days, a bigger variation of temperature values is included in the training set.
This supports the general idea that when we train the models with more data, our predictions will improve.
With a baseline, trained on 360 training days, the Retailer is able to estimate, with reasonable accuracy, the energy savings resulting from its energy policies.
Moreover, he can compare the energy savings to the investment made for the measure.
This has obvious advantages for the retailer.

In summary, the results of this study show that we have been able to create reliable energy baselines using off-the-shelf data science technologies.
Moreover, we found a way to create them based on short term historical data.

\section*{Acknowledgments}
This work is financed by the ERDF – European Regional Development Fund through the COMPETE Programme (operational programme for competitiveness) and by National Funds through the FCT – Fundação para a Ciência e a Tecnologia (Portuguese Foundation for Science and Technology) within project 3GEnergy (AE2016-0286).

\section*{References}
\bibliographystyle{abbrv}
\bibliography{bibio}

\end{document}